\def\ls{\lower4pt\hbox{${\buildrel < \over \sim}$}}
\def\gs{\lower4pt\hbox{${\buildrel > \over \sim}$}}

\documentclass[12pt, preprint]{aastex}

\slugcomment{Accepted for publication in {\it The Astrophysical Journal}}

\shorttitle{Variability of 3C~279}
\shortauthors{M. B\"ottcher \& D. Principe}

\begin{document}

\title{The Optical Variability of the Quasar 3C~279: The Signature
of a Decelerating Jet?}

\author{M. B\"ottcher and D. Principe\altaffilmark{1}}

\altaffiltext{1}{Astrophysical Institute, Department of Physics and Astronomy, \\
Clippinger 339, Ohio University, Athens, OH 45701, USA}

\begin{abstract}
A recent optical monitoring campaign on the prominent quasar
3C~279 revealed at least one period of a remarkably clean exponential
decay of monochromatic (BVRI) fluxes with time, with a time constant
of $\tau_d = 12.8$~d, over about 14 days. This is clearly too
long to be associated with radiative cooling. Here we propose that
this may be the signature of deceleration of the synchrotron emitting 
jet component. We develop a model analogous to the relativistic blast 
wave model for gamma-ray bursts, including radiative energy losses and 
radiation drag, to simulate the deceleration of a relativistically
moving plasmoid in the moderately dense AGN environment. Synchrotron,
SSC and external Compton emission are evaluated self-consistently. 
We show that the observed optical light curve decay can be successfully
reproduced with this model.  

The decelerating plasmoid model predicts a delayed X-ray flare, about
2 -- 3 weeks after the onset of the quasi-exponential light curve decay 
in the optical. A robust prediction of this model, which can be tested 
with Fermi and simultaneous optical monitoring, is that the peak in 
the $\gamma$-ray light curve at $\sim 100$~MeV is expected to be delayed by
a few days with respect to the onset of the optical decay, while the VHE 
$\gamma$-rays are expected to track the optical light curve closely
with a delay of at most a few hours. 

\end{abstract}


\section{Introduction}

The quasar 3C~279 ($z = 0.536$) is one of the best-observed flat 
spectrum radio quasars, in part because of its prominent 
$\gamma$-ray flare shortly after the launch of the {\it Compton
Gamma-Ray Observatory (CGRO)} in 1991. It has been 
persistently detected by the {\it Energetic Gamma-ray 
Experiment Telescope (EGRET) on board {\it CGRO}} 
each time it was observed, even in its very low quiescent 
states, e.g., in the winter of 1992 -- 1993, and is 
known to vary in $\gamma$-ray flux by roughly two orders of 
magnitude \citep{maraschi94,wehrle98}. It has been monitored
intensively at radio, optical, and more recently also X-ray
frequencies, and has been the subject of intensive multiwavelength
campaigns \citep[e.g.,][]{maraschi94,hartman96,wehrle98}. 
The most recent multiwavelength campaign on 3C~279 included
a Whole Earth Blazar Telescope (WEBT) campaign in the spring
of 2006 \citep{boettcher07b}. During this campaign, the source 
was overall in a high optical state, with $R \sim 14.0$ -- $14.5$. 
However, the light curves showed an extraordinary feature: An 
unusually clean, quasi-exponential decay of the BVRI fluxes with 
a time scale of $\tau_d = 12.8$~d, extended over about 2 weeks.
This paper aims at a theoretical interpretation of this 
extraordinary light curve feature.

Flat-spectrum radio quasars (FSRQs) and BL~Lac objects are 
active galactic nuclei (AGNs) commonly unified in the class 
of blazars. They exhibit some of the most violent high-energy
phenomena observed in AGNs to date. Their spectral energy
distributions (SEDs) are characterized by non-thermal continuum 
spectra with a broad low-frequency component in the radio -- UV 
or X-ray frequency range and a high-frequency component from
X-rays to $\gamma$-rays. 
In the framework of relativistic jet models, the low-frequency (radio
-- optical/UV) emission from blazars is interpreted as synchrotron
emission from nonthermal electrons in a relativistic jet. The
high-frequency (X-ray -- $\gamma$-ray) emission could either be
produced via Compton upscattering of low frequency radiation by the
same electrons responsible for the synchrotron emission \citep[leptonic
jet models; for a recent review see, e.g.,][]{boettcher07a}, or 
due to hadronic processes initiated by relativistic protons 
co-accelerated with the electrons \citep[hadronic models, for 
a recent discussion see, e.g.,][]{muecke01,muecke03}. 
Several authors have modeled broadband SEDs of 3C~279 in various
states \citep[e.g.,][]{bednarek98,sikora01,hartman01,moderski03}.
A consistent picture emerges that the X-ray -- soft $\gamma$-ray 
portion of the SED might be dominated by 
synchrotron self-Compton (SSC)
emission, while the 
{\it EGRET} emission might require an additional component, most 
likely external Compton emission. 

Standard leptonic models of blazar emission generally assume that 
a relativistic plasmoid containing ultrarelativistic nonthermal
electrons moves with constant bulk Lorentz factor $\Gamma$ along
a jet, directed at a small angle with respect to our line of sight.
However, for several blazars, in particular high-frequency peaked
BL Lac objects detected at $> 100$~GeV $\gamma$-rays, such models 
sometimes require unexpectedly large bulk Lorentz factors ($\Gamma 
\gtrsim 50$) and accordingly small viewing angles in order to explain 
their SEDs and variability \citep{bfr08,gt08,finke08}. 
Such large Lorentz factors and small viewing angles pose serious
problems for AGN unification schemes, according to which FR I radio
galaxies are believed to be the unbeamed equivalents of BL~Lac
objects. A possible solution to this dilemma might lie in the
deceleration of the emission region \citep{gk03a,gk03b} from 
sub-pc scales, at which the optical -- X-ray -- $\gamma$-ray 
emission is produced, towards pc and kpc scales, which can be 
resolved with VLBA / VLBI techniques. At those scales, superluminal
speeds of individual jet components of $\beta_{\rm app} \lesssim 10$
are characteristically observed in most cases, providing an estimate 
of the Lorentz factor of jet components at those scales of $\Gamma 
\sim 10$. In fact, extreme deceleration of a radio-emitting plasmoid 
(component C3) in the jet of 3C~279 may already habe been directly 
observed in space VLBI monitoring observations \citep{piner00},
although the identification of this component over multiple
observing epochs with different instruments/arrays is highly 
uncertain.

In this paper, we propose a model analogous to the relativistic
blast wave model which has successfully predicted and explained
the smooth, self-similar light curves of X-ray and optical afterglows
of $\gamma$-ray bursts \citep{pr93,mr97,cd99}. We adapt this model
for the specific situation in blazars. In particular, we include
self-consistently radiative losses and radiation drag from
Comptonization of external radiation fields. A similar study,
with emphasis on the details of the isotropization of particle
distributions in the plasmoid and on spectral features from
various leptonic and hadronic processes, has been performed
by \cite{ps00}, who find good agreement of their results with 
characteristic SEDs of blazars. Here, we adopt a simplified 
description of the particle dynamics and radiation processes, 
and focus on the expected monochromatic light curves dominated 
by the plasmoid deceleration. We review the observational
motivation from 3C~279 in \S \ref{motivation}, describe the 
model for the plasmoid dynamics in \S \ref{model}, and outline 
our treatment of radiation processes in \S \ref{radiation}. As a
test of our numerical simulations, we develop an analytical 
solution to the plasmoid dynamics and light curves in the
self-similar deceleration phase in \S \ref{deceleration}.
In \S \ref{results} we present results of our simulations
and fits to the observed exponential flux decay of 3C~279 
in January 2006. We summarize in \S \ref{summary}.

Throughout this paper, we refer to $\alpha$ as the energy 
spectral index, $F_{\nu}$~[Jy]~$\propto \nu^{-\alpha}$. A 
cosmology with $\Omega_m = 0.3$, $\Omega_{\Lambda} = 0.7$, 
and $H_0 = 70$~km~s$^{-1}$~Mpc$^{-1}$ is used. In this cosmology,
and using the redshift of $z = 0.536$, the luminosity distance 
of 3C~279 is $d_L = 3.08$~Gpc. 

\begin{figure}[ht]
\plotone{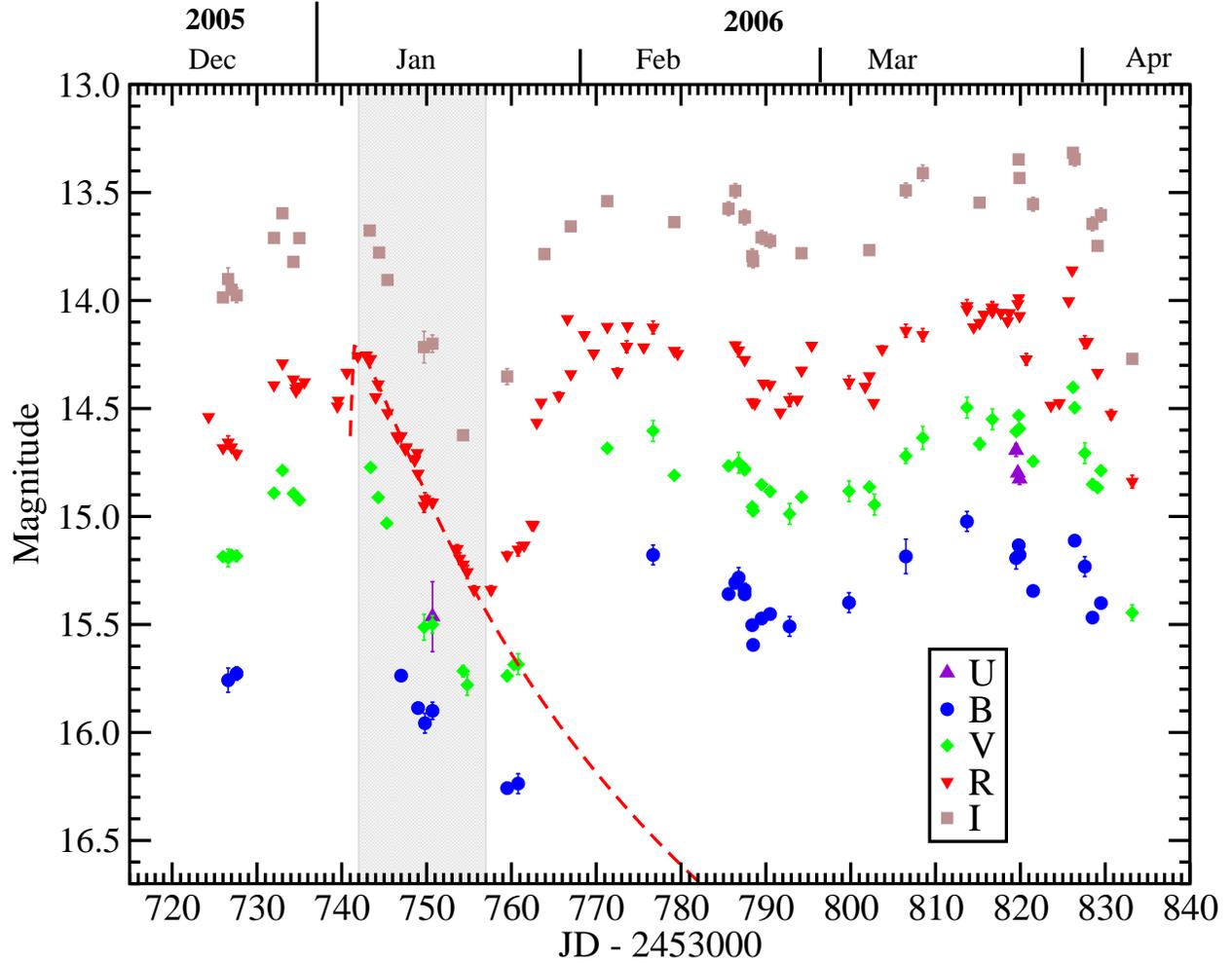}
\caption{Light curves of 3C~279 in various optical bands during the
spring of 2006. The dotted red line indicates our model fit, as 
discussed in the following sections, to the R-band light curve during 
the quasi-exponential decay around January 15, 2006}
\label{lightcurves}
\end{figure}

\section{\label{motivation}Observational Motivation}

3C~279 was observed in a WEBT campaign at radio, near-IR, and optical 
frequencies, throughout the spring of 2006. Details of the observations,
data analysis, and implications of the optical variability patterns
observed during that campaign have been published in \cite{boettcher07b}.
Fig. \ref{lightcurves} shows the optical light curves of 3C~279 during 
spring 2006. The light curves exhibit an extraordinarily clean quasi-exponential
decay with a characteristic time scale of $\tau_d \sim 12.8$~days around
JD 2453743 -- JD 2453760. This light curve feature can not be interpreted 
as the signature of radiative cooling since the synchrotron cooling time 
scale for electrons emitting synchrotron radiation in the optical R band 
is

\begin{equation}
\tau_{\rm sy}^{\rm obs} \sim 3 \times 10^4 \, B_G^{-1/2} D_1^{-1/2} \; {\rm s}
\label{tausy}
\end{equation}
where $B_G$ is the magnetic field in Gauss and $D_1$ is the Doppler
factor in units of 10. This
is of the order of at most a few hours for typical values of the 
magnetic field strength expected in quasars ($B \sim 1$~G). Setting the 
synchrotron cooling time scale equal to the observed exponential decay time 
scale, would require a magnetic field of $B \sim 7 \times 10^{-4} \, D_1^{-1}$~G, 
which is about three orders of magnitude lower than usually inferred for quasar 
jets. We therefore favor a model in which the light curve decay is associated
with the dynamics of the emission region rather than microscopic processes. 
We note that similar quasi-exponential decays have also been observed 
in 3C~279 repeatedly in the 2007 observing season \citep{larionov08}.

\section{\label{model}Model of a decelerating jet}

Our treatment of a decelerating jet is borrowed from the blast wave model
of gamma-ray bursts. For details see, e.g., \cite{cd99}. We assume a plasmoid
moving ballistically with initial mass $M_0$ and bulk Lorentz factor
$\Gamma_0$ along the jet. Let $M$ be the relativistic mass of the the
plasmoid in the rest-frame of the plasmoid, then the momentum $P$ of the
plasmoid in the stationary AGN frame is $P = \beta \, \Gamma \, M \, c$,
where $\beta$ is the normalized velocity $v/c$ corresponding to the bulk
Lorentz factor $\Gamma$. If the plasmoid radiates isotropically in its rest
frame, the equation of motion of the plasmoid can be derived from momentum
conservation, $dP/dt = 0$. However, in the case of a quasar, a substantial 
contribution to the (bolometrically dominant) $\gamma$-ray emission results 
from Compton upscattering of external radiation fields (EC = External Compton), 
and a significant transfer of plasmoid momentum to Compton-scattered external 
radiation (``Compton drag'') has to be taken into account. We can therefore 
write

\begin{equation}
{dP \over dt} = \left( {dP \over dt} \right)_{\rm EC} = {c \, M 
\dot\Gamma \over \beta} + \Gamma \beta \, {\dot M} \, c
\label{dPdt}
\end{equation}
For large enough $\Gamma$, most of the EC radiation will be beamed into a
narrow cone of solid angle $\Omega \sim 1/\Gamma^2$, and we can write
the momentum transfer to EC radiation as

\begin{equation}
\left( {dP \over dt} \right)_{\rm EC} = - {1 \over c} \int\limits_{4\pi} 
{dL/d\Omega} \cos\theta \, d\Omega \approx {\Gamma^2 \over 4 \pi \, c} 
\, \dot{E'}_{\rm EC}
\label{dPdtEC}
\end{equation}

where $\dot{E'}_{\rm EC}$ is the internal energy loss due to EC radiation
in the co-moving frame. For the purpose of an approximate, quantitative
analysis to extract the salient spectral and light curve features of this
model, we assume that all Compton scattering occurs in the Thomson regime 
so that

\begin{equation}
\dot{E'}_{\rm EC} = - {1 \over \Gamma} \, {4 \over 3} c \, \sigma_T \, 
{u'}_{\rm ext} \int\limits_{1}^{\infty} N_e (\gamma) \, \gamma^2 \, d\gamma.
\label{EdotEC}
\end{equation}
The factor $1/\Gamma$ in Eq. (\ref{EdotEC}) stems from the fact that 
$\dot{E'}$ constitutes a derivative with respect to time in the stationary
AGN frame, and $dt' = dt/\Gamma$. We can also use Eq. (\ref{dPdtEC}) 
for a rough estimate of the magnitude of the radiation drag force, assuming
that the observed $\gamma$-ray emission results from Compton scattering of
an isotropic radiation field. Then, 

\begin{equation}
\left({dP \over dt}\right)_{\rm drag} \approx - {(1 + z) \, d_L^2 \, 
\nu F_{\nu}^{\rm pk, EC} \over \Gamma^2 c} \sim 4.3 \times 10^{33} 
\, f_{13} \, \Gamma_1^{-2} \; {\rm dyne}
\label{radiationdrag}
\end{equation}
where $f_{13} = \nu F_{\nu}^{\rm pk, EC} / (10^{13}$~Jy~Hz) and $\Gamma_1
= \Gamma/10$. This may be compared to an estimate for the Compton rocket
effect due to radiation from the accretion disk. If we approximate the
accretion disk radiation as a point source (the most optimistic estimate)
with luminosity $L_D \equiv 10^{46} \, L_{46}$~erg~s$^{-1}$, impinging from
behind on an emission region of radius $R_b \equiv 10^{16} \, R_{16}$~cm,
located at a distance $r \equiv 0.1 \, r_{-1}$~pc from the accretion disk,
the force on the plasmoid due to the Compton rocket effect can be estimated
as

\begin{equation}
\left({dP \over dt}\right)_{\rm acc} \approx {L_D \, R_b^2 \over 16 \, 
r^2 \, c} \approx 2.3 \times 10^{31} \, L_{46} \, R_{16}^2 \, r_{-1}^{-2} 
\; {\rm dyne}.
\label{comptonrocket}
\end{equation}
Thus, for standard parameters with $r \gtrsim 10^{-2}$~pc and $L_D \lesssim
10^{46}$~erg~s$^{-1}$, the Compton rocket effect may safely be neglected in 
our calculations.

The accumulation and radiative loss of relativistic mass $\dot M$ can be 
calculated as

\begin{equation}
\dot M = A(r) \rho(r) \Gamma(r) \, {dr/dt} + {1 \over c^2} \dot{E'}_{\rm rad} 
\label{Mdot}
\end{equation}
where $dr/dt = \beta c$, $A(r) = \pi \, R_b^2 (r)$ is the cross section of the 
jet, $\rho(r)$ is the density of external material being swept up by the plasmoid, 
and

\begin{equation}
\dot{E'}_{\rm rad} = - {1 \over \Gamma} \, {4 \over 3} c \, \sigma_T \, u' 
\int\limits_{1}^{\infty} N_e (\gamma) \, \gamma^2 \, d\gamma 
\label{Edotrad}
\end{equation}
Here, $u'$ is the sum of the energy densities, $u'_{\rm B} + u'_{\rm ext} 
+ u'_{\rm sy}$, and, again, we have assumed Compton scattering to be dominated 
in the Thomson regime. 

Derivatives with respect to time can be converted to derivatives
with respect to distance $r$ from the central engine, yielding

\begin{equation}
{d\Gamma \over dr} = - {\Gamma(r) \beta^2(r) \over M(r)} \,
{dM \over dr} + {\Gamma^2 \, \dot{E'}_{\rm EC} \over 4 \pi \, M \, c^3}
\label{dGdr}
\end{equation}
and
\begin{equation}
{dM \over dr} = A(r) \, \rho(r) \, \Gamma (r) + {\dot{E'}_{\rm rad} \over 
\Gamma(r) \beta(r) \, c^3}
\label{dMdr}
\end{equation}

We assume that a fraction of the swept-up electrons will be instantaneously
accelerated into a power-law, which can be described by an injection function
of the form

\begin{equation}
Q(\gamma) = Q_0 \, \gamma^{-q} \, H(\gamma_{\rm min}, \gamma, \gamma_{\rm max}),
\label{Qg}
\end{equation}
where $H(x_0, x, x_1) = 1$ for $x_0 < x < x_1$ and 0 otherwise. The low- and 
high-energy cutoffs of the electron injection function are given by

\begin{equation}
\gamma_{\rm min} = {\epsilon_e \over \xi_e} \, \left( {q - 2 \over q - 1} 
\right) \, {m_p \over m_e} \, \Gamma
\label{gmin}
\end{equation}
where $\epsilon_e$ is the fraction of swept-up power that is transferred to
relativistic electrons, 
$\xi_e$ is the fraction of swept-up electrons which
is accelerated to ultrarelativistic energies. The maximum Lorentz 
factor can be estimated by balancing the fastest conceivable acceleration
time scale (the Larmor time scale) with the synchrotron loss time scale:

\begin{equation}
\gamma_{\max} \sim 4.7 \times 10^7 \, B_G^{-1/2}
\label{gmax}
\end{equation}

where $B_G$ is the magnetic field in Gauss.
The magnetic field may be parameterized in terms of a fraction $e_B$ 
of the swept-up energy transferred to magnetic-field energy density:

\begin{equation}
B (r) = \sqrt{ 32 \, \pi \, e_B \, \rho(r) \, c^2 \,} \, \Gamma (r)
\label{B}
\end{equation}

where an additional factor of 4 has been introduced to account for 
the compression of the pre-shock material by the strong shock.
We can find the normalization $Q_0$ of the
electron injection function through

\begin{equation}
Q_0(r) = \xi_e \, {A(r) \, \Gamma(r) \, \beta(r) \, \rho(r) \, c \over m_p}
\, {1 - q \over \gamma_{\rm max}^{1 - q} - \gamma_{\rm min}^{1 - q}}.
\label{Q0}
\end{equation}

Eq. (\ref{tausy}) indicates that the characteristic radiative cooling time 
scales of particles emitting synchrotron radiation at optical or higher
frequencies are likely to be much shorter than the dynamical time scale
of the system. Therefore, the balance of relativistic particle acceleration,
cooling, and escape on a time scale

\begin{equation}
\tau'_{\rm esc} = \eta \, {R_b \over c}
\label{tauesc}
\end{equation}
yields a critical electron energy $\gamma_c$ beyond which particles effectively
radiate their energy away on a time scale shorter than the escape time scale:

\begin{equation}
\gamma_c = {3 \, m_e c^2 \over 4 \, \sigma_T \, u' \, \eta \, R_b}
\label{gammac}
\end{equation}
The resulting quasi-equilibrium electron energy distribution will be a broken
powerlaw with parameters depending on whether $\gamma_c > \gamma_{\rm min}$
(slow-cooling regime) or $\gamma_c < \gamma_{\rm min}$ (fast-cooling regime). 
Re-writing the radiative cooling rate as $\dot\gamma \equiv - \nu_0 \, \gamma^2$
with $\nu_0 = (4/3) \, c \, \sigma_T \, u' / (m_e c^2)$ we find for
the slow-cooling regime

\begin{equation}
N_{\rm sc} (\gamma, t) \approx 
\cases{
{Q_0 \, t_{\rm esc} \over (q - 1)} \, \gamma^{-q} 
     & for $\gamma_{\rm min} < \gamma < \gamma_c$ \cr\cr
{Q_0 \over \nu_0 \, (q - 1)} \, \gamma^{-(q + 1)} 
     & for $\gamma_c < \gamma < \gamma_{\rm max}$ \cr
}
\label{Nsc}
\end{equation}
In the fast-cooling regime, we have

\begin{equation}
N_{\rm fc} (\gamma, t) \approx 
\cases{
{Q_0 \over \nu_0 \, (q - 1)} \, \gamma_{\rm min}^{1 - q} \, \gamma^{-2} 
     & for $\gamma_c < \gamma < \gamma_{\rm min}$ \cr\cr
{Q_0 \over \nu_0 \, (q - 1)} \, \gamma^{-(q + 1)} 
     & for $\gamma_{\rm min} < \gamma < \gamma_{\rm max}$ \cr
}
\label{Nfc}
\end{equation}

The apparent large break in the spectral index around the synchrotron peak
in the SED of 3C~279 indicates that this peak can not be attributed to a 
cooling break, but rather to a large value of $\gamma_{\rm min}$, and the 
system is in the fast-cooling regime \citep{boettcher08}. 
Eqs. (\ref{gmin}) and (\ref{gammac}) naturally yield values that
support the assumption of the fast-cooling regime for plausible parameter
values:

\begin{equation}
\gamma_{\rm min} \sim 600 {\epsilon_{e, -1} \over \xi_e} \Gamma_1
\label{gminnum}
\end{equation}
where $\epsilon_{e, -1} \equiv \epsilon_e / 0.1$, while

\begin{equation}
\gamma_c \lesssim 30 \, \Gamma_1^{-2} \, R_{16}^{-1}
\label{gcnum}
\end{equation}
where we have only taken into account Compton cooling on the external
radiation field for a characteristic value of $u'_{\rm ext}$ as given 
in Eq. (\ref{ublr}). Consequently, $\gamma_{\rm min} > \gamma_{\rm c}$
as long as (formally) $\Gamma > 0.5 \, R_{16}^{-1} \, \xi_e / \epsilon_{e, -1}$,
which will be the case for all plausible parameter values, even when 
the plasmoid becomes non-relativistic.

\section{\label{radiation}Radiation}

The relativistic electrons described by the distribution functions
(\ref{Nsc}) and (\ref{Nfc}), will emit synchrotron, synchrotron-self-Compton
(SSC) and EC radiation. For simplicity, we express the synchrotron
emissivity using a $\delta$ function approximation:

\begin{equation}
\nu F_{\nu}^{\rm sy} (\nu_{\rm obs}) = {D^4 \, \epsilon' \, c \, \sigma_T \, 
{u'}_B \over 6 \, \pi \, d_L^2 \, \epsilon_B \, (1 + z)} \left( {\epsilon' 
\over \epsilon_B} \right)^{1/2} \, N_e \left( \sqrt{\epsilon' \over \epsilon_B} 
\right)
\label{Fsy}
\end{equation}
where $\epsilon_B = B/B_{\rm crit}$ with $B_{\rm crit} = 4.414\times 
10^{14}$~G, and $\epsilon' = ([1 + z] / D) \, h \nu_{\rm obs} / (m_e c^2)$.
The corresponding photon number density is

\begin{equation}
{n'}_{\rm sy} ({\epsilon'}_{\rm sy}) = {3 \, \sigma_T \, {u'}_B \over 8 \,
\pi \, \epsilon_B \, {\epsilon'}_{\rm sy} \, m_e c^2 \, R_b^2} \, 
\left( {{\epsilon'}_{\rm sy} \over \epsilon_B} \right)^{1/2} \, 
N_e \left( \sqrt{{\epsilon'}_{\rm sy} 
\over \epsilon_B} \right).
\label{nsy}
\end{equation}
Writing the electron distribution (\ref{Nsc}) or (\ref{Nfc}) as 

\begin{equation}
N_e (\gamma) = N_0 \, \cases{
\left( {\gamma \over \gamma_b} \right)^{-p_1} & for $\gamma_1 < \gamma < \gamma_b$ \cr\cr
\left( {\gamma \over \gamma_b} \right)^{-p_2} & for $\gamma_b < \gamma < \gamma_2$ \cr
}
\label{Nnew}
\end{equation}
with $(\gamma_1, \gamma_b, \gamma_2) = (\gamma_{\rm min}, \gamma_c, \gamma_{\rm max})$
and $(p_1, p_2) = (q, q+1)$ for the slow-cooling regime and $(\gamma_1, \gamma_b, 
\gamma_2) = (\gamma_c, \gamma_{\rm min}, \gamma_{\rm max})$ and $(p_1, p_2) =
(2, q+1)$ for the fast-cooling regime, we can express the photon energy density 
as

\begin{equation}
{u'}_{\rm syn} = {16 \over 9} \, \sigma_T \, {u'}_B { R_B \, N_0 \over V_B}
\left( \gamma_b^{p_1} {\gamma_b^{3 - p_1} - \gamma_1^{3 - p_1} \over 3 - p_1}
+ \gamma_b^{p_2} {\gamma_2^{3 - p_2} - \gamma_b^{3 - p_2} \over 3 - p_2}
\right).
\label{usy}
\end{equation}

The photon number density (\ref{nsy}) can be used in the \cite{jones68} formula
to evaluate the SSC flux:

\begin{equation}
\nu F_{\nu}^{\rm SSC} (\nu_{\rm obs}) = {D^4 \, {\epsilon'}^2 \, m_e c^2 \
\over 4 \pi \, d_L^2 \, (1 + z)} \int\limits_1^{\infty} d\gamma \, 
N_e (\gamma) \, \int\limits_0^{\infty} d{\epsilon'}_{\rm sy} \, {n'}_{\rm sy} 
({\epsilon'}_{\rm sy}) \, g(\epsilon', {\epsilon'}_{\rm sy}, \gamma)
\label{nFnssc}
\end{equation}
with
\begin{equation}
g(\epsilon', {\epsilon'}_{\rm sy}, \gamma) = {c \, \pi \, r_e^2 \over 2 \, 
\gamma^4 {\epsilon'}_{\rm sy}}
\left( {4 \, \gamma^2 \, \epsilon' \over {\epsilon'}_{\rm sy}} - 1 \right) 
\;\;\;\;\;\; {\rm if}
\;\;\;\;\;\; {{\epsilon'}_{\rm sy} \over 4 \, \gamma^2} \le \epsilon' \le 
{\epsilon'}_{\rm sy},
\label{g1}
\end{equation}
and
$$
g(\epsilon', {\epsilon'}_{\rm sy}, \gamma) = {2 \, c \, \pi \, r_e^2 \over 
\gamma^2 {\epsilon'}_{\rm sy}} \left( 2 \, q \, \ln q + (1 + 2 q) (1 - q) + 
{(4 \, {\epsilon'}_{\rm sy} \, \gamma \, q)^2 \over (1 + 4 \, {\epsilon'}_{\rm sy} 
\, \gamma \, q)} \, {(1 - q) \over 
2} \right) 
$$
\begin{equation}
{\rm if} \;\;\;\;\;\;\;\;\;\; {\epsilon'}_{\rm sy} \le \epsilon' \le 
{4 \, {\epsilon'}_{\rm sy} \, \gamma^2 \over 1 + 4 \, {\epsilon'}_{\rm sy} 
\, \gamma}
\label{g2}
\end{equation}
where
\begin{equation}
q = {\epsilon' \over 4 \, {\epsilon'}_{\rm sy} \, \gamma^2 \left( 1 - 
{\epsilon' \over \gamma} \right)}.
\label{q}
\end{equation}

We evaluate the external-Compton photon spectrum with a $\delta$ function
approximation for the external radiation field, 

\begin{equation}
{n'}_{\rm ext} (\epsilon, \Omega) \approx {{u'}_{\rm ext}
\over {\epsilon'}_{\rm ext} \, m_e c^2} \, \delta(\epsilon - {\epsilon'}_{\rm ext})
\, \delta(\mu' + 1)
\label{next}
\end{equation}
where $\mu' = \cos{\theta'}_{\rm ext}$ refers to the angle of incidence of the
external photons with respect to the jet axis, and ${\epsilon'}_{\rm ext} =
\Gamma \epsilon_{\rm ext}$. The Compton cross section is also approximated by 
a $\delta$ function,

\begin{equation}
{d^2 \sigma_C \over d\Omega' \, d\epsilon'} \approx \sigma_T \, \delta (\Omega'
- {\Omega'}_e) \, \delta (\epsilon' - \gamma^2 \, {\epsilon'}_{\rm ext} \,
[1 - \beta \mu]).
\label{sigmac}
\end{equation}
With these simplifications, the EC flux can be calculated as

\begin{equation}
\nu F_{\nu}^{\rm EC} (\nu_{\rm obs}) = {D^4 \, c \, \sigma_T \, {u'}_{\rm ext} 
\over 8 \, \pi \, d_L^2 \, (1 + z)} \, \left( {{\epsilon'} \over 
{\epsilon'}_{\rm ext}} \right)^{3/2} \, \sqrt{1 + \mu_{\rm obs}} \, 
N_e \left( \sqrt{\epsilon' \over {\epsilon'}_{\rm ext} \, (1 + \mu_{\rm obs})} 
\right).
\label{nFnec}
\end{equation}

\section{\label{deceleration}Asymptotic Behaviour in the Deceleration Phase}

As in the well-known case of expanding blast waves in Gamma-Ray Bursts
(GRBs), the plasmoid in a decelerating jet (with constant cross section
$A$) is starting out in a coasting phase, in which the initial mass $M_0$
greatly exceeds the swept-up relativistic mass in the co-moving frame.
During this phase, the effect of the inertia of the swept-up mass is
negligible, and the Lorentz factor remains roughly constant. This phase
is followed by the deceleration phase. In the asymptotic limit of that
phase, the initial mass of the plasmoid becomes negligible. 

Although we have properly included the effect of radiative cooling and
radiation drag, in most cases the fraction of swept-up energy which is
transferred to ultrarelativistic electrons and can therefore be radiated
away efficiently, will be small. Therefore, we can approximate the equation
of motion of the plasmoid by an adiabatic solution, as

\begin{equation}
{d\Gamma \over dr} \approx {- \Gamma^2 \, A \, \rho_{\rm ext} \over M}
\label{dGdrdecel}
\end{equation}
and
\begin{equation}
M(r) \approx A \, \rho_{\rm ext} \int\limits_{r_0}^r \Gamma (r') \, dr'
\label{Mdecel}
\end{equation}

where we have, for simplicity, assumed a constant external density
$\rho(r) \equiv \rho_{\rm ext} =$~const. This system has a self-similar 
solution of the form

\begin{equation}
\Gamma(r) = \Gamma_0 \, \left( {r \over r_0} \right)^{-1/2}.
\label{Gammadecel}
\end{equation}
Assuming for the purpose of an analytic estimate that we are looking
right down the jet ($\theta_{\rm obs} = 0$), the observer's time
as a function of Lorentz factor can be expressed as

\begin{equation}
t_{\rm obs} = t_0 + \int\limits_{r_0}^r { (1 - \beta(r') \cos\theta_{\rm obs})
\over \beta(r') c} \, dr' \approx {1 \over 2 \, c} \int\limits_{r_0}^r
{dr' \over \Gamma^2 (r')} \approx {r^2 \over 4 \, r_0 \, \Gamma_0^2 \, c}.
\label{tobsdecel}
\end{equation}
This yields a solution for the Lorentz factor as a function of observer's
time:

\begin{equation}
\Gamma(t_{\rm obs}) \approx \sqrt{\Gamma_0 \over 2} \, \left( {r_0 \over
c \, t_{\rm obs}} \right)^{1/4} \propto t_{\rm obs}^{-1/4}.
\label{Gamma_tobs}
\end{equation}

The observed steep spectral index of the optical synchrotron emission
from 3C~279 ($\alpha \sim 1.7$), indicates $p \sim 4.4$. This, in turn,
signifies that the system is in the fast cooling regime since otherwise 
a cooling break would not produce a $\nu F_{\nu}$ peak at the synchrotron 
frequency corresponding to $\gamma_b$. Furthermore,
electrons synchrotron radiating at optical frequencies are most likely
beyond the break energy, i.e., $\gamma > \gamma_{\rm min}$. For the 
prediction of synchrotron light curves we may therefore use Eq. \ref{Fsy}
together with the lower branch of Eq. \ref{Nfc}. For $\gamma_{\rm max}
\gg \gamma_{\rm min}$, the coefficient $Q_0$ in Eq. \ref{Nfc} may be
approximated as

\begin{equation}
Q_0(r) \approx {\xi_e \, A \, \Gamma(r) \, \rho_{\rm ext} \, c \over m_p} 
\, (q - 1) \, \gamma_{\rm min}^{q - 1}(r) \propto \Gamma^q.
\label{Q0decel}
\end{equation}
The cooling coefficient $\nu_0$ is expected to be dominated by synchrotron
and/or external Compton cooling. With the magnetic-field scaling from
Eq. \ref{B}, both $u'_{\rm ext}$ and $u'_B$ carry a dependence $\propto
\Gamma^2$. Thus, we find for the normalization of the ultrarelativistic
particle population:

\begin{equation}
N_e (\gamma, t_{\rm obs}) \propto \Gamma^{q - 2} \, \gamma^{-(q + 1)}
\propto t_{\rm obs}^{(2 - q)/4} \, \gamma^{-(q + 1)}.
\label{Nedecel}
\end{equation}

In order to use Eq. \ref{Fsy} for a light curve estimate, we assume,
again, for simplicity, $\theta_{\rm obs} = 0$ and therefore $D \approx
2 \, \Gamma$. Consequently, the characteristic electron energy 
$\gamma = \sqrt{\epsilon' / \epsilon_B} \propto \Gamma^{-1}$.
This yields an expected light curve decay in the fast-cooling 
synchrotron regime of

\begin{equation}
\nu F_{\nu}^{\rm sy} (\nu_{\rm obs}, t_{\rm obs}) \propto 
\nu_{\rm obs}^{(2 - q)/2} \, \Gamma^{2 \, (1 + q)} \propto
\nu_{\rm obs}^{(2 - q)/2} \, t_{\rm obs}^{- (1 + q)/2}.
\label{Fsylightcurve}
\end{equation}
In particular, for an injection index of $q = 3.4$, as inferred from the 
optical spectral index, a light curve of $F_{\nu} \propto t_{\rm obs}^{-2.2}$
is expected. However, it should be pointed out that this 
can only be considered an upper limit to the steepness of the decay. 
Any non-zero observing angle will flatten the decay of the light curve 
as it introduces a shallower decay of the Doppler factor $D$ with 
decreasing $\Gamma$ and therefore with time. In any case, by the 
time the plasmoid is in the self-similar deceleration phase, the 
monochromatic flux has already decreased by about an order of magnitude 
from its initial peak value, and is likely to be overwhelmed by other 
emission components in the jet. We do therefore not expect to observe
the limiting deceleration case directly. We only developed this analytical
case to demonstrate the agreement of our numerical simulations with the 
analytical expectation in the following section.

\begin{figure}[ht]
\plotone{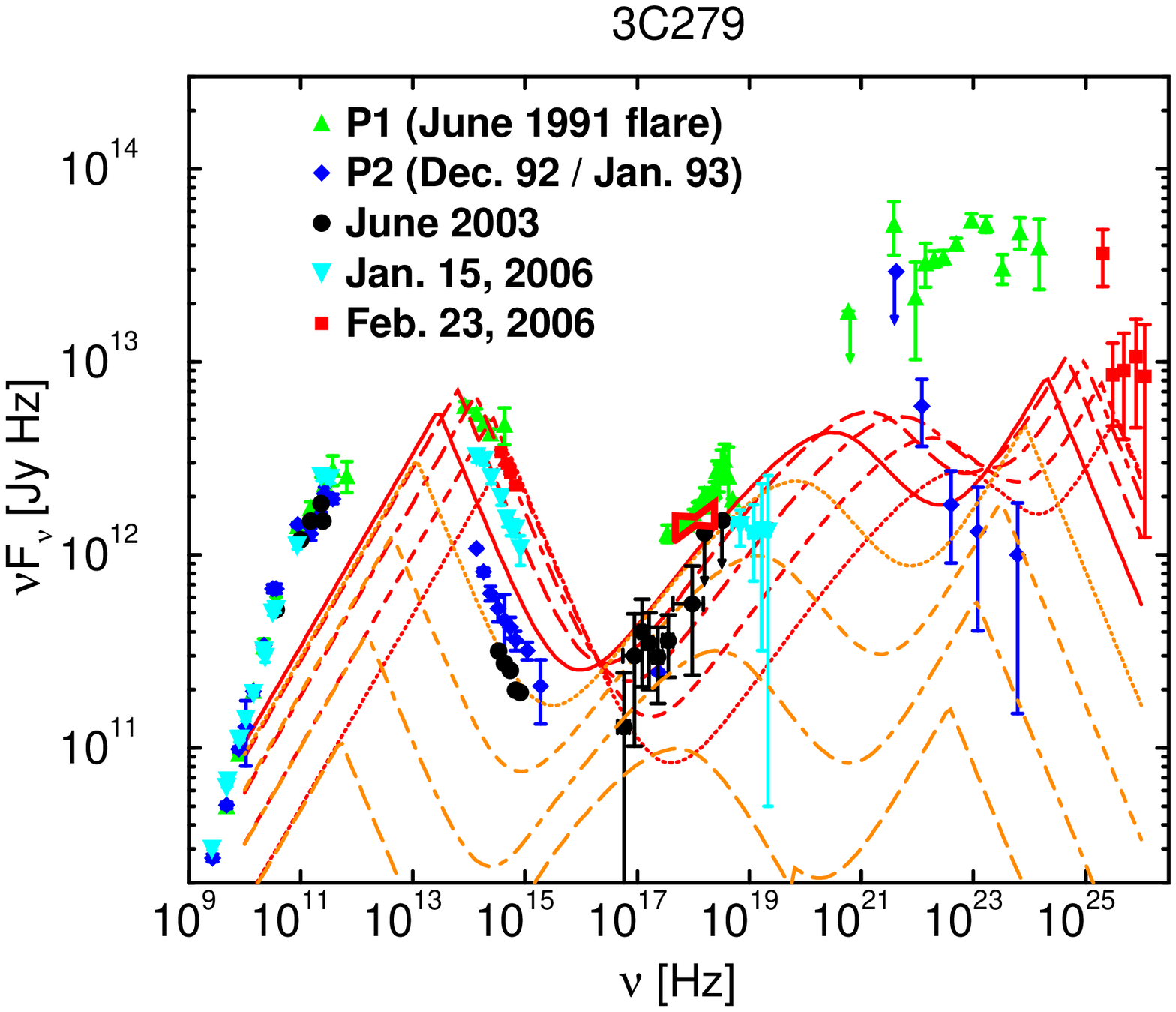}
\caption{Simulated snap-shot SEDs from our decelerating plasmoid model,
compared to various observed SEDs of 3C~279 in various observing epochs
and activity states. The time sequence goes from red - dotted $\to$ dashed
$\to$ dot-dashed $\to$ long-dashed $\to$ solid $\to$ orange - dotted $\to$ 
dashed $\to$ dot-dashed $\to$ long-dashed. This simulation provides a good
fit to the optical light curve decay of 3C~279 around January 15, 2006 
(see Fig. \ref{lcfits}). }
\label{spsequence}
\end{figure}

\section{\label{results}Numerical Results}

In order to highlight the salient features of our model, we assume a simple
cylindrical jet geometry with a constant cross section of the jet, $A \equiv
\pi R_b^2$, as well as a homogeneous external medium with density $\rho (r)
\equiv \rho_{\rm ext}$. 
We achieved good fits to the observed optical light
curves and overall SED shape of 3C~279 with the parameters listed in Table 
\ref{parameters}. Our choice of the external density corresponds to
a number density of $n_{\rm ext} = 100$~cm$^{-3}$. This is at least about
an order of magnitude lower than typical particle densities in the broad
line regions of quasars. This is quite reasonable since we expect that the
jet trajectory is already partially evacuated from previous ejection events.

Fig. \ref{spsequence} compares snap-shot SEDs of our simulation to various
observed broadband SEDs of 3C~279. It can be seen that the optical continuum
spectra in various intensity states are well represented by the model, the
X-ray spectral slope corresponds to the characteristically observed shape,
and for much of the plasmoid evolution, the simulated X-ray flux is in the
range of observed values. The $\gamma$-ray flux represented by our simulation
corresponds to a medium to low state of 3C~279. Note that this work is not
intended to attempt a model interpretation of the very-high-energy $\gamma$-ray
flux detected by MAGIC on Feb. 23, 2006 \citep{albert08}, in particular 
since we did not take into account any effects of intrinsic or intergalactic 
$\gamma\gamma$ absorption. For model implications of this high-energy detection 
see \cite{boettcher08}.

\begin{figure}[ht]
\plotone{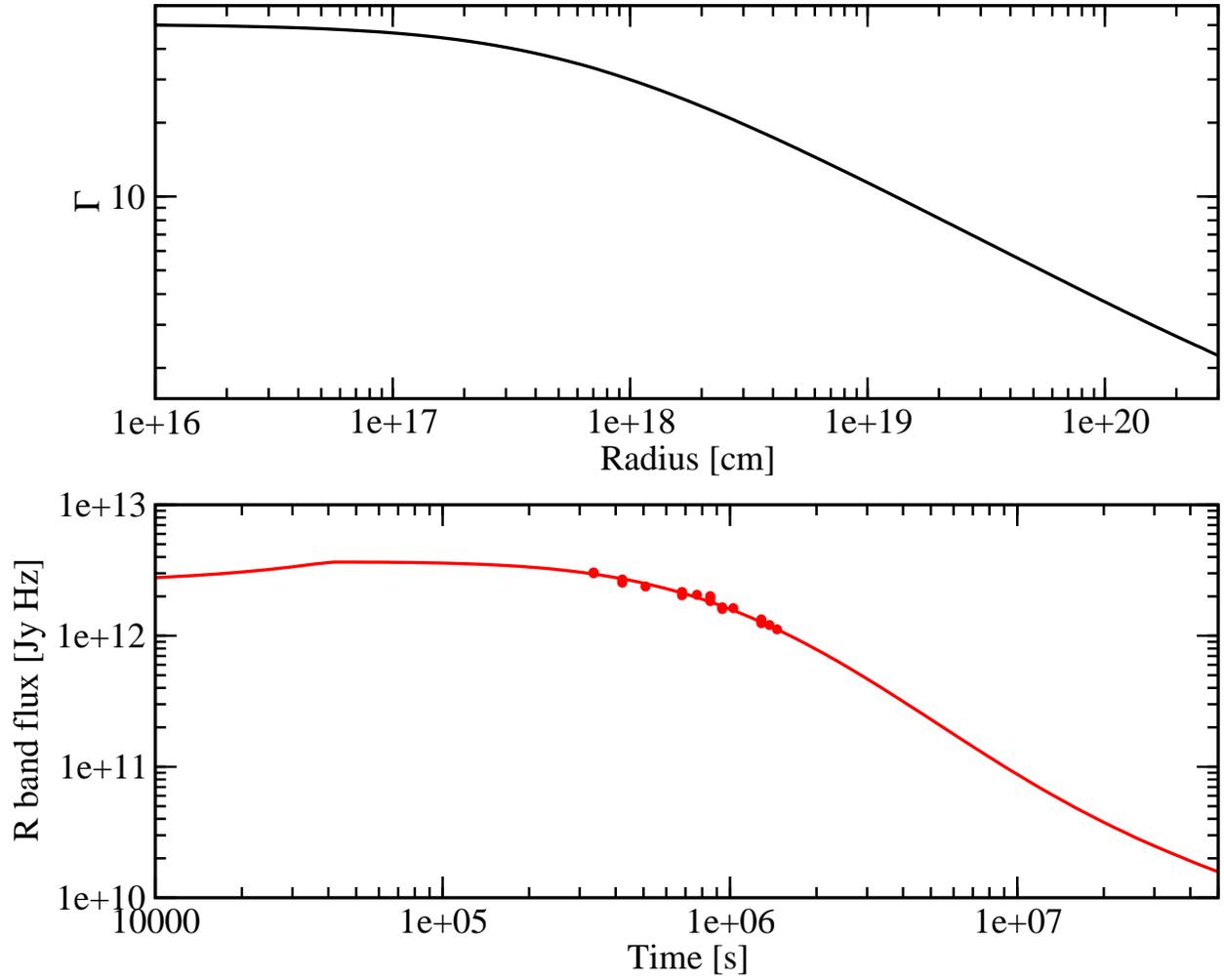}
\caption{Top: Evolution of Lorentz factor $\Gamma$ as a function of distance
$r$ from the central engine for the simulation illustrated in Fig. 
\ref{spsequence}. Bottom: R-band light curve from this simulation.}
\label{complete}
\end{figure}

The upper panel in figure \ref{complete} illustrates 
the evolution of the plasmoid bulk Lorentz factor $\Gamma$ as a function of 
distance from the central engine, obtained by numerically solving the coupled 
system of Eqs. (\ref{dGdr}) and (\ref{dMdr}). 

As in the well-known case of the blast-wave model for gamma-ray bursts,
the plasmoid evolves through a coasting phase with approximately constant
Lorentz factor $\Gamma_0$. Around the characteristic deceleration radius
$r_d$, at with the swept-up relativistic mass equals the mass of the initial
ejecta $M_0$, the evolution makes a gradual transition into the asymptotic
self-similar deceleration phase with $\Gamma (r) \propto r^{-1/2}$, treated 
analytically in \S \ref{deceleration}. For the input parameters listed in 
Table \ref{parameters}, this occurs at

\begin{equation}
r_d = {M_0 \over \Gamma_0 \, \pi \, R_b^2 \, \rho_{\rm ext}} \approx 1.1 
\times 10^{18} \; {\rm cm}.
\label{rd}
\end{equation}

Note, however, that substantial deceleration happens long before $r_d$ is 
reached. The lower panel of Fig. \ref{complete} shows the resulting R-band light
curve over the entire evolution of the plasmoid, down to a mildly relativistic
speed. Again, we see the gradual turnover to the self-similar deceleration phase 
with the expected light curve decay as $F_{\rm sy} \propto t_{\rm obs}^{-2.2}$ 
(see \S \ref{deceleration}). During the transition from the coasting to the 
self-similar deceleration phase, the light curve is reasonably well represented 
by an exponential decay. 

\begin{figure}[ht]
\plotone{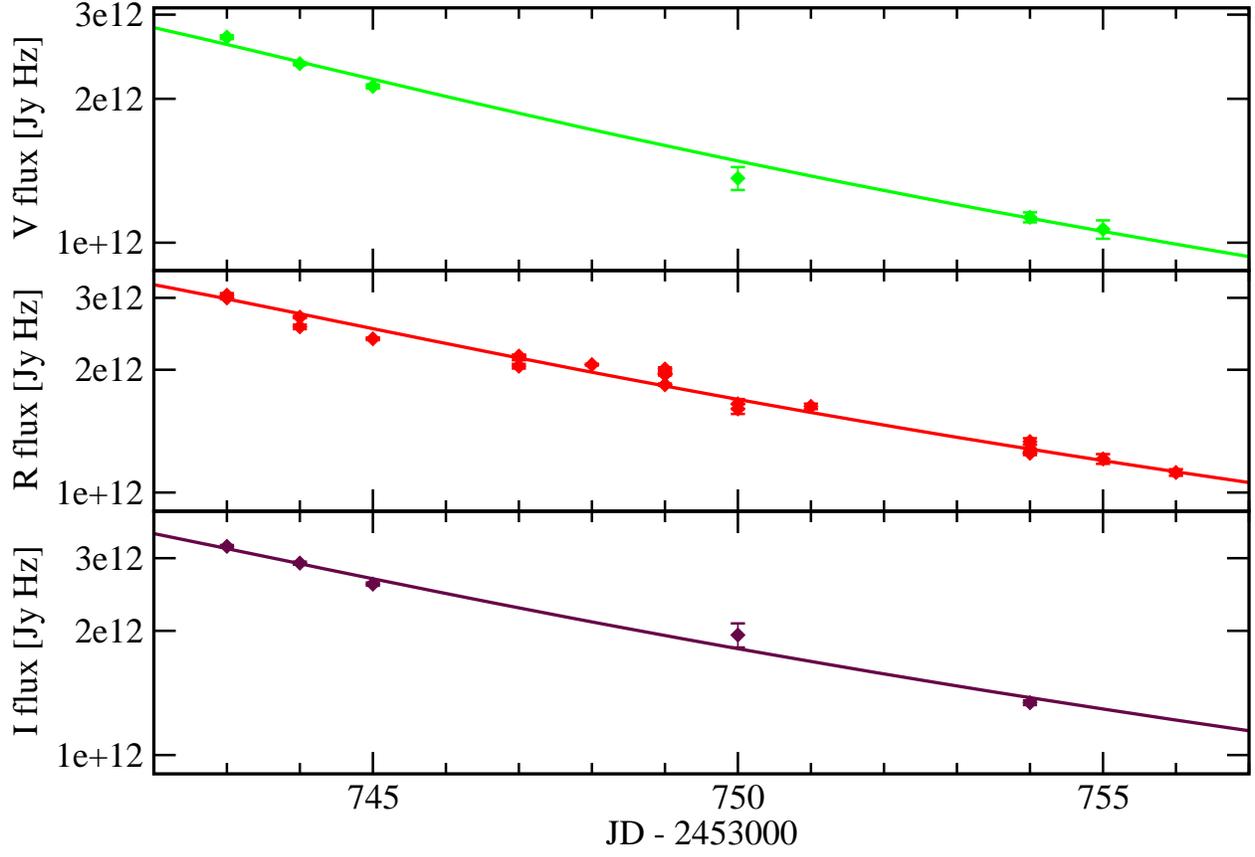}
\caption{Fit of our decelerating-jet model to the V, R, and I band fluxes
of 3C~279. }
\label{lcfits}
\end{figure}

The detailed fits to the V, R, and I band light curves of 3C~279 during the
quasi-exponential decay around January 15, 2006, are shown in Fig. \ref{lcfits}. 
The B and U band light curves were rather poorly sampled during this time
period so that the (equally good) fits to those light curves does not provide
substantial additional information. 

\begin{figure}[ht]
\plotone{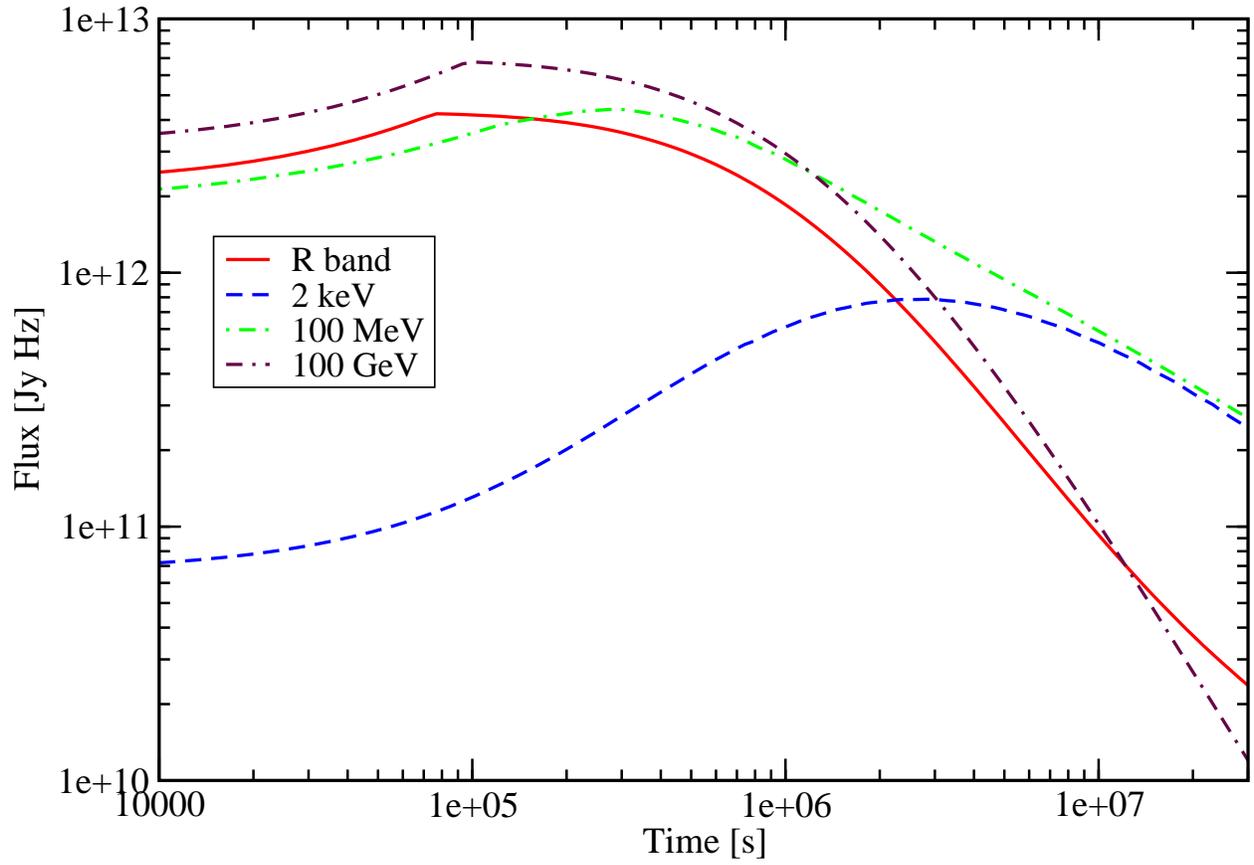}
\caption{Predicted X-ray and $\gamma$-ray light curves for parameters from
our fit to the optical light curves of 3C~279. }
\label{Xgammalc}
\end{figure}

Using characteristic parameters for the accretion disk luminosity, $L_D \equiv 
10^{45} \, L_{45}$~ergs~s$^{-1}$ and the Thomson depth $\tau_{\rm T, BLR} \equiv
10^{-1} \, \tau_{-1}$ and radius $R_{\rm BLR} \equiv 0.1 \, R_{-1}$~pc of
the BLR of 3C~279, we estimate an external radiation energy density of

\begin{equation}
u_{\rm ext} \approx {L_D \, \tau_{\rm T, BLR} \over 4 \, \pi \, 
R_{\rm BLR}^2 \, c} \sim 3 \times 10^{-3} \, L_{45} \, \tau_{-1} \, 
R_{-1}^{-2} \; {\rm ergs \; cm}^{-3}
\label{ublr}
\end{equation}
and $\epsilon_{\rm ext} \sim 10^{-5}$ \citep[see, e.g.,][]{boettcher08}, 
which we use for our calculation of the EC $\gamma$-ray emission component 
and the associated radiation drag term. The predicted light curves of the 
combined SSC + EC emissions at X-rays and $\gamma$-rays are compared to
the R-band light curves in Fig. \ref{Xgammalc}. 

The most remarkable feature of the predicted light curves is a delayed 
X-ray outburst about 2 -- 3 weeks after the onset of the optical decline. 
This is a consequence of the shift of the SSC peak towards lower
frequencies as the blob decelerates, which follows from the dependence
$\nu_{\rm pk, SSC} \propto B^2 \, \gamma_{\rm min}^4 \propto \Gamma^6$.
This SSC peak frequency decrease is much more rapid than the decrease
of the SSC peak flux. This spectral evolution of the SSC component is 
illustrated in Fig. \ref{spsequence}.
In fact, RXTE monitoring of the source \citep[see][]{boettcher08}
did detect a transition of 3C~279 from a quiescent state throughout the first
half of January 2006, to a very active, high X ray flux state with a substantial
X-ray outburst around Feb. 5, 2006. It is tempting to postulate that this may
have been the X-ray signature of the plasmoid deceleration observed during
the optical decay around January 15.

A prediction of this scenario is that the $\sim 100$~MeV $\gamma$-ray light
curve peaks a few days after the optical before it begins its quasi-exponential 
decay. This is a consequence of the fact that this energy range consists of
comparable contributions from the SSC and the EC radiation components. As the
SSC component shifts towards lower energies during the early deceleration, the
EC component just moves into the 100 MeV range, before it also decays away to
lower peak energies and lower flux levels. The very-high-energy ($\gtrsim 100$~GeV)
light curve is expected to follow closely the optical one, with only a small
delay, which critically depends on the initial bulk Lorentz factor and the 
mean photon energy of the external photon field.

We caution that any predictions concerning absolute flux levels at X-rays 
and $\gamma$-rays are very model-parameter dependent and vary substantially 
for different choices of initial mass, initial bulk Lorentz factor, initial 
pladmoid radius, external matter density, external radiation field, etc. 
Furthermore, we have not included any effects of intrinsic \citep[in particular, 
in the radiation field of the BLR, see, e.g.,][]{dp03,reimer07,sb08}, nor 
intergalactic $\gamma\gamma$  absorption. However, while this may significantly 
affect the overall $\gamma$-ray flux level, the light curve features discussed 
above are predominantly a consequence of the plasmoid dynamics, which are 
dictated by the observed optical light curves, and are therefore robust 
predictions of the decelerating-jet model.

\section{\label{summary}Summary and Conclusions}

Motivated by an extraordinarily clean quasi-exponential decay of the
V, R, and I band light curves of 3C~279 over a period of about 2 weeks 
during a recent WEBT campaign, we proposed a model of a decelerating
plasmoid in the jet of this quasar. We take into account self-consistently
the inertia of swept-up mass as the plasmoid propagates through the gas 
of the AGN environment, as well as radiation drag and radiative cooling.
We have demonstrated that, similar to the relativistic blast wave model
for GRBs, the plasmoid makes a transition from a coasting phase with
approximately constant Lorentz factor, to a self-similar phase. In the
case of a homogeneous external medium and a cylindrical jet, the self-similar
deceleration phase is described by $\Gamma (r) \propto r^{-1/2} \propto
t_{\rm obs}^{-1/4}$. The resulting optical synchrotron light curves are
well approximated by quasi-exponential decays during the transition from
the coasting to the self-similar decay phase. In the asymptotic limit
of the self-similar decay phase, the synchrotron light curves at frequencies
corresponding to electron energies above the cooling break of the electron 
spectrum follow a behaviour of $\nu F_{\nu} (\nu_{\rm obs} \, t_{\rm obs})
\propto \nu_{\rm obs}^{(2 - q)/2} \, t_{\rm obs}^{-(1 + q)/2}$, where
$q$ is the injection spectral index of ultrarelativistic electrons in
the plasmoid. 

We note that the choice of a conical jet instead of a cylindrical 
one would recover the isotropic blast wave model for gamma-ray bursts 
\citep[e.g.,][]{cd99}: It would lead to a scaling of the plasmoid surface
as $A \propto r^2$. In the self-similar deceleration phase, this recovers
the well-known adiabatic blast wave solution with $\Gamma (r) \propto
r^{-3/2} \propto t_{\rm obs}^{-8/3}$. This case describes a much faster
deceleration of the plasmoid and therefore substantially steeper light
curves than the cylindrical jet geometry assumed here.

We have demonstrated that this model can adequately reproduce the observed
optical light curves of 3C~279 during the $\sim 2$ week long quasi-exponential 
decay phase in January 2006. This model predicts a delayed (SSC-dominated)
X-ray outburst about 2 -- 3 weeks after the onset of the optical decay.
We speculate that the X-ray flare around Feb. 5, 2006, detected by {\it RXTE}
monitoring, may have been the X-ray signature of the plasmoid deceleration
seen earlier in the optical bands. 

A robust prediction of the decelerating plasmoid model which can be tested 
with Fermi and simultaneous optical monitoring is that the peak in the 
$\gamma$-ray light curve at $\sim 100$~MeV is expected to be delayed by
a few days with respect to the onset of a quasi-exponential light curve
decay in the optical, while the VHE $\gamma$-rays are expected to track
the optical light curve closely. 

The quasar 3C~279 is one of the most active blazars known. The ejection of 
$\gamma$-ray emitting plasmoids from the nucleus of 3C~279 might be a frequent 
event. We do therefore expect to observe the quasi-exponential decay phase
following such ejection events only in a temporary quiescent phase, in which
it is not overwhelmed by subsequent ejection events. This might be the reason
why these quasi-exponential decays in the light curves of this blazar are not 
more frequently observed. In addition to one or two occurrences during the
2006 WEBT campaign, several more quasi-exponentially decaying light curve
segments can also be identified in the 2007 WEBT campaign data \citep{larionov08},
indicating that these are rare, but not unique events.

\acknowledgments
We thank the anonymous referee for a very constructive report. 
We thank Matthew G. Baring for stimulating discussions. This work
was supported by NASA through XMM-Newton Guest Observer Grant
NNX08AD67G.

\begin{deluxetable}{ccccc}
\tabletypesize{\scriptsize}
\tablecaption{Parameters of our plasmoid evolution simulation providing a fit
to the V, R, and I band light curves of 3C~279 around January 15, 2006.}
\tablewidth{0pt}
\tablehead{
\colhead{Parameter} & \colhead{Symbol} & \colhead{Value} }
\startdata
Initial Lorentz factor 		& $\Gamma_0$		& $50$ \\
External matter density 	& $n_{\rm ext} = \rho_{\rm ext} / m_p$	& $100$~cm$^{-3}$ \\
Plasmoid radius 		& $R_b$			& $3 \times 10^{16}$~cm \\
Initial mass			& $M_0$			& $2.6 \times 10^{31}$~g \\
Electron injection index 	& $q$			& $3.4$ \\
B-field equipartition parameter & $e_B$			& $10^{-3}$ \\
Electron acceleration efficiency & $\epsilon_e$		& $0.1$ \\
Relativistic electron fraction 	& $\xi_e$		& $0.5$ \\
Electron escape time scale parameter & $\eta$		& $10$ \\
External radiation energy density & $u_{\rm ext}$	& $3 \times 10^{-3}$~erg~cm$^{-3}$ \\
Observing angle 		& $\theta_{\rm obs}$	& $5^o$ \\
\enddata
\label{parameters}
\end{deluxetable}

\end{document}